\pdfoutput=1

\documentclass{svjour3}                     
\smartqed  
\usepackage{amssymb}

\usepackage{pdfpages}

\begin{document}

\title{Classical model of confinement
}


\author{Yu. P. Goncharov         \and
        N. E. Firsova 
}


\institute{Yu. P. Goncharov \at
Theoretical Group, Experimental Physics Department, 
State Polytechnical University, 
Sankt-Petersburg 195251, Russia\\
              \email{ygonch@chern.hop.stu.neva.ru}           
           \and
           N. E. Firsova \at
Institute of the Mechanical Engineering Problems, 
Russian Academy of Sciences, 
Sankt-Petersburg 199178, Russia\\
\email{nef2@mail.ru}
}

\date{Received: date / Accepted: date}

\maketitle

\begin{abstract}
The confinement mechanism proposed earlier and then applied successfully to 
meson spectroscopy by one of the authors is interpreted in classical terms. 
For this aim the unique solution of the Maxwell 
equations, an analog of the corresponding unique solution of the 
SU(3)-Yang-Mills equations describing linear confinement in quantum 
chromodynamics, is used. Motion of a charged particle is studied in the field 
representing magnetic part of the mentioned solution and it is 
shown that one deals with the full classical confinement of the charged 
particle in such a field: under any initial conditions the particle motion is 
accomplished within a finite region of space so that the particle trajectory 
is near magnetic field lines while the latter are compact manifolds (circles). 
An asymptotical expansion for the trajectory form in the strong field limit is 
adduced. The possible application of the obtained results in thermonuclear 
plasma physics is also shortly outlined.  


\end{abstract}

\section{Introduction}
In Refs. \cite{{Gon01},{Gon051},{Gon052}} for the Dirac-Yang-Mills 
system derived from 
QCD-Lagrangian an unique family of compatible 
nonperturbative solutions was found and explored, which could pretend to 
decsribing confinement of two quarks. 
The successful applications of the family to the description of both the heavy 
quarkonia spectra \cite{{Gon03a},{Gon03b},{Gon04},{Gon08a}} and a number of 
properties of pions, kaons, $\eta$ and 
$\eta^\prime$-meson \cite{{Gon06},{Gon07a},{Gon07b},{Gon08},{Gon08b}} showed 
that the confinement mechanism is qualitatively the same for both light mesons 
and heavy quarkonia and it is mainly governed by the magnetic colour field 
linear in $r$ (distance between quarks) which represents a part of the mentioned 
unique family of solutions and, in its turn, the part is a solution of the 
SU(3)-Yang-Mills equations. 

As has been emphasized, however, as far back as in 
Refs. \cite{{Gon051},{Gon052}}, the similar unique confining solutions exist for any 
compact semisimple and non-semisimple Lie groups, in particular, for 
SU$(N)$-groups with $N\ge2$ and for U$(N)$-groups with $N\ge1$, i.e. it holds 
true also for classical electrodynamics with group U$(1)$ and Maxwell equations. 
Under this situation, as was pointed out in 
Refs. \cite{{Gon051},{Gon052}}, 
there is an interesting possibility of indirect experimental verification 
of the confinement mechanism under discussion. Indeed the confining solutions 
of Maxwell equations for classical electrodynamics point out 
the confinement phase could be in electrodynamics as well. Though 
there exist no elementary charged particles generating a constant magnetic 
field linear in $r$, the distance from particle, after all, if it could 
generate this elecromagnetic field configuration in laboratory then one might 
study motion of the charged particles in that field. The confining properties 
of the mentioned field should be displayed at classical level too but the exact 
behaviour of particles in this field requires certain analysis of the corresponding 
classical equations of motion.

The aim of the present paper  
is to some degree to realize the above program on studying motion of 
the charged particles in the mentioned confining electromagnetic field.   

Section 2 contains preliminaries necessary to pose the problem: information on 
the confining solutions of the Yang-Mills and Maxwell equations and on the 
miscellaneous forms of the motion equations for a charged particle in the 
confining magnetic field when considering it with using different curvilinear 
coordinates. Section 3 is devoted to the general conclusions of a 
qualitative character concerning behaviour of a charged particle in the 
magnetic field under discussion. In the strong field limit Section 4 gives 
asymptotical expansions for the spherical coordinates of a particle when its 
moving in the field under consideration while Section 5 contains numerical 
estimates and Section 6 is devoted to discussion and concluding remarks. 
                      
Appendix A is devoted to the formulation of vector analysis on a 
region $\Omega$ in ${\mathbb R}^3$ which is most convenient, especially while 
working with using the arbitrary curvilinear coordinates so the mentioned 
formulation is employed throughout the paper. At last, Appendix B supplements 
Section 2 with a proof of the uniqueness theorem from that Section in the case 
of U(1)-group (Maxwell equations).

Also throughout the paper we employ the Heaviside-Lorentz system of units 
with $\hbar=c=1$ and also with the Boltzmann constant $k=1$, unless explicitly 
stated otherwise. When calculating we apply the 
relations $1\ {\rm GeV^{-1}}\approx0.1973269679\ {\rm fm}\>$,
$1\ {\rm s^{-1}}\approx0.658211915\times10^{-24}\ {\rm GeV}\>$, 
$1\ {\rm V/m}\approx0.2309956375\times 10^{-23}\ {\rm GeV}^2$, 
$1\ {\rm T}=4\pi\times10^{-7} {\rm H/m}\times1\ {\rm A/m}
\approx0.6925075988\times 10^{-15}\ {\rm GeV}^2 $. 
\section{Preliminaries}
\subsection{The confining solutions of SU(3)-Yang-Mills and Maxwell 
equations}
As was mentioned above, our study is motivated by the confinement mechanism 
proposed earlier by one of the authors and based on the unique family 
of compatible nonperturbative solutions for 
the Dirac-Yang-Mills system (derived from QCD-Lagrangian) studied at the whole 
length in Refs. \cite{{Gon01},{Gon051},{Gon052}}.  

One part of the mentioned family is presented by the unique nonperturbative 
confining solution of the SU(3)-Yang-Mills 
equations for gluonic field $A=A_\mu dx^\mu=
A^a_\mu \lambda_adx^\mu$ ($\lambda_a$ are the 
known Gell-Mann matrices, $\mu=t,r,\vartheta,\varphi$, $a=1,...,8$). To specify 
the question, let us note that in general the Yang-Mills equations on a 
manifold $M$ can be written as
$$d\ast F= g(\ast F\wedge A - A\wedge\ast F) \>,\eqno(1)$$ 
where the curvature matrix (field strentgh)
$F=dA+gA\wedge A= F^a_{\mu\nu}\lambda_adx^\mu\wedge dx^\nu$ with exterior 
differential $d$ and the Cartan's (exterior) product $\wedge$, while $\ast$ 
means the Hodge star operator conforming to a metric on manifold under 
consideration, $g$ is a gauge coupling constant.

The most important case of $M$ is Minkowski spacetime and we 
are interested in the confining solutions $A$ of the SU(3)-Yang-Mills 
equations. The confining solutions were defined in Ref. \cite{Gon01} as the 
spherically symmetric solutions of the Yang-Mills 
equations (1) containing only the components of the 
SU($3$)-field which are Coulomb-like or linear in $r$. Additionally 
we impose the Lorentz condition on the sought solutions. 
The latter condition is necessary for 
quantizing the gauge fields consistently within the framework of perturbation 
theory (see, e. g. Ref. \cite{Ryd85}), so we should impose the given condition 
that can be written
in the form ${\rm div}(A)=0$, where the divergence of the Lie algebra valued
1-form $A=A_\mu dx^\mu=A^a_\mu \lambda_adx^\mu$ is defined by the relation 
(see, e. g., Refs. \cite{{Bes87},{Po1}})
$${\rm div}(A)=\frac{1}{\sqrt{\delta}}\partial_\mu(\sqrt{\delta}g^{\mu\nu}
A_\nu)\>.\eqno(2)$$
It should be emphasized that, from the physical point of view, the Lorentz 
condition reflects the fact of transversality for gluons that arise as quanta 
of SU(3)-Yang-Mills field when quantizing the latter (see, e. g., 
Ref. \cite{Ryd85}).
  
Under the circumstances, the unique nonperturbative confining solution of 
the SU(3)-Yang-Mills equations
looks as follows 
$$ A^3_t+\frac{1}{\sqrt{3}}A^8_t =-\frac{a_1}{r}+A_1 \>,
 -A^3_t+\frac{1}{\sqrt{3}}A^8_t=-\frac{a_2}{r}+A_2\>,$$
$$-\frac{2}{\sqrt{3}}A^8_t=\frac{a_1+a_2}{r}-(A_1+A_2)\>, $$
$$ A^3_\varphi+\frac{1}{\sqrt{3}}A^8_\varphi=
b_1r+B_1 \>,
 -A^3_\varphi+\frac{1}{\sqrt{3}}A^8_\varphi=
b_2r+B_2\>,$$
$$-\frac{2}{\sqrt{3}}A^8_\varphi=
-(b_1+b_2)r-(B_1+B_2)\> \eqno(3)$$
with the real constants $a_j, A_j, b_j, B_j$ parametrizing the family. 
As has been repeatedly discussed by us earlier (see, e. g., 
Refs. \cite{{Gon051},{Gon052}} and below), from the above form it is clear that 
the solution (3) is a configuration describing the electric Coulomb-like colour 
field (components $A^{3,8}_t$) and the magnetic colour field linear in $r$ 
(components $A^{3,8}_\varphi$) and we wrote down
the solution (3) in the combinations that are just 
needed further to insert into the corresponding Dirac equation (for more 
details see Refs. \cite{{Gon01},{Gon051},{Gon052}}).
 
The word {\em unique} should be understood in the strict mathematical sense. 
In fact in Ref. \cite{Gon051} the following theorem was proved (see also 
Appendix B):

{\em The unique exact spherically symmetric (nonperturbative) confining 
solutions (depending only on $r$ and $r^{-1}$) of SU(3)-Yang-Mills 
equations in Minkowski spacetime consist of the family of (3)}.

It should be noted that solution (3) was found early in 
Ref. \cite{Gon01} but its uniqueness was proved just in Ref. \cite{Gon051} 
(see also Ref. \cite{Gon052}). Besides, in Ref. \cite{Gon051} it was shown 
that the above unique confining solutions (3) 
satisfy the so-called Wilson confinement criterion \cite{{Wil},{Ban}}. Up to now 
nobody contested this result so if we want to describe interaction between 
quarks by spherically symmetric confining SU(3)-fields then they can be only 
those from the above theorem.

Now one should say that the similar unique confining solutions exist for all 
semisimple and non-semisimple compact Lie groups, in particular, for SU($N$) 
with $N\ge2$ and 
U($N$) with $N\ge1$ \cite{{Gon051},{Gon052}}. Explicit form of solutions, 
e.g., for SU($N$) with $N=2,4$ can be found in Ref.\cite{Gon052} but it 
should be emphasized that components linear in $r$ always represent the 
magnetic (colour) field in all the mentioned solutions. Within the present 
paper we are especially interested in the U(1)-case (electrodynamics) and 
a proof of the above 
uniqueness theorem for that situation is adduced in Appendix B for inquiring. 

Under this situation the Yang-Mills equations (1) turn into the second pair of 
Maxwell equations 
$$d\ast F= 0 \eqno(4)$$ 
with $F=dA$, $A=A_\mu dx^\mu$. 
As is discussed in Appendix B, in the spherically symmetric case the equations (4) 
are equivalent to 
$$\partial_r(r^2\partial_rA_t)=0,\>\partial^2_rA_\varphi=0\>,\eqno(5)$$
with $A_t=A_t(r)$, $A_\varphi=A_\varphi(r)$ and we write down the unique 
solutions of (5) as 
$$ A_t =\frac{a}{r}+A \>, A_\varphi=br+B \>\eqno(6)$$
with some constants $a, b, A, B$ parametrizing solutions. 
 
To interpret solutions (6) in the more habitual physical terms let us pass on
to Cartesian coordinates employing the relations 
$$\varphi=\arctan(y/x),\>
d\varphi=\frac{\partial\varphi}{\partial x}dx+
\frac{\partial\varphi}{\partial y}dy\> \eqno(7)$$
which entails
$${\bf A}=A_\varphi d\varphi=(br+B)d\varphi=
-\frac{(br+B)y}{x^2+y^2}dx+\frac{(br+B)x}{x^2+y^2}dy\>
\eqno(8)$$
and we conclude that the solutions (6) describe the combination
of the electric Coulomb field with potential $\Phi=A_t$ and the constant 
magnetic field with the vector-potential (8) which can be written as 
(using isomorphism $dx\Longleftrightarrow{\bf i}$, 
$dy\Longleftrightarrow{\bf j}$, $dz\Longleftrightarrow{\bf k}$, see Appendix A)
$${\bf A}=A_x{\bf i}+A_y{\bf j}+A_z{\bf k}=
-\frac{(br+B)y}{x^2+y^2}{\bf i}+\frac{(br+B)x}{x^2+y^2}{\bf j}\>,
\eqno(9)$$
which is {\it linear} in $r$ in spherical coordinates. Let us compute 
3-dimensional divergence ${\mathrm{div}}{\bf A}$ with the help 
of 3-dimensional Hodge star operator (see Appendix A). In spherical 
coordinates we have [see (A.12) and (A.7)]
$${\mathrm{div}}{\bf A}=\ast(d\ast {\bf A})=\ast d\ast[(br+B)d\varphi]=
\ast d\left(\frac{br+B}{\sin{\vartheta}}dr\wedge d\vartheta\right)=0\>. $$

Then eqs. (4) in 
Cartesian coordinates take the form
$$\Delta\Phi=0,\> {\rm rot\,rot}{\bf A}= {\mathrm{grad}}\,{\mathrm{div}}{\bf A}-
\Delta{\bf A}=-\Delta{\bf A}=0\> \eqno(10)$$
with the Laplace operator $\Delta=\partial_x^2+\partial_y^2+\partial_z^2$.
Also it is easy to check that the solution under consideration satisfies the 
4-dimensional Lorentz condition ${\rm div}(A)=0$, where 
4-dimensional divergence is defined by (2). 

Finally, as is shown in Appendix A, the corresponding strength of magnetic field 
is $${\bf H}={\mathrm {rot}}\,{\bf A}=\ast(d{\bf A})=
-\frac{b}{\sin{\vartheta}}d\vartheta= -\frac{b}{r}\left(\frac{xz}{x^2+y^2}
dx+\frac{yz}{x^2+y^2}dy-dz\right)$$
$$\Longleftrightarrow-\frac{b}{r}\left(\frac{xz}{x^2+y^2}{\bf i}+
\frac{yz}{x^2+y^2}{\bf j}-{\bf k}\right)\>,\eqno(11) $$
respectively, in spherical and Cartesian coordinates. 
\subsection{Singularities of solutions}
As is seen from (8) and (11), the magnetic field under exploration has the 
singularities on the $z$-axis so its formal mathematical definition domain is 
the manifold ${\mathbb{R}}^3\backslash \{z\}$ with the $z$-axis discarded 
rather than the manifold ${\Bbb{R}}^3$. Singularities of such a kind are of 
mathematical nature and appear when trying to write a {\em concrete 
macroscopic physical field} in an analytical form. Physical origin of the given 
singularities is that some sources generating the field should be present on 
$z$-axis. Another matter is that the field under consideration may probably be 
modelled by miscellaneous ways. For the sake of completeness, one of possible 
physical realization will be considered in Sec. 6. In theoretical considerations 
within classical approach one should segregate from a concrete realization of 
one or another macroscopic electromagnetic field and consider them to be given 
on their natural mathematical definition domains. 

At the quantum level, however, treatment of singularities may be different 
from classical one. In particular, in the case of gluonic field (3) 
the problem of singularity along $z$-axis of magnetic part for 
solution (3) can be resolved by that quarks may emit gluons outside of some 
cone $\vartheta=\vartheta_0$ so singularity along $z$-axis plays no role 
(for more details see Ref. \cite{Gon08b} and estimates for $\vartheta_0$ in 
pions and kaons therein).

\subsection{Equations of motion for a charged particle in the confining 
magnetic field}
As was mentioned in Section 1, at quantum level the confinement of quarks is 
basically governed by the magnetic (colour) part (linear in $r$) of solution 
(3), as has been discussed in 
Refs. 
\cite{{Gon03a},{Gon03b},{Gon04},{Gon08a},{Gon06},{Gon07a},{Gon07b},{Gon08},{Gon08b}}. 
In the present paper we would like, at classical level, to explore the behaviour 
of a charged 
particle moving in the confining magnetic field (11). Accordingly, we need 
to study classical equations of motion for such a particle. As is known 
(see, e.g., Ref. \cite{LL}), those equations are obtained from Lagrangian
$$L=-m\sqrt{1-v^2}+q{\bf A}{\bf v}, \eqno(12)$$
where $q$ and $m$ are, respectively, charge and mass of a particle while 
the form of both the velocity square $v^2=g^{\mu\nu}v_\mu v_\nu$ and   
the scalar product ${\bf A}{\bf v}= g^{\mu\nu}A_\mu v_\nu$ depends on choice of 
curvilinear coordinates. Then the sought equations are derived from (12) 
according to the standard prescription of Lagrange approach as
$$\frac{d}{dt}\left(\frac{\partial L}{\partial \dot Q_i}\right)-
\left(\frac{\partial L}{\partial Q_i}\right)=0,\>i=1,2,3\>, \eqno(13)$$                     
where $Q_i$ are the chosen coordinates and the dot signifies differentiation 
with respect to $t$. For our purposes the equations of motion will be useful 
in both spherical and Cartesian coordinates. One can note that $v^2$ is 
conserved \cite{LL} when moving in a constant magnetic field, i.e., the 
full energy $E=m/\sqrt{1-v^2}$ of (relativistic) particle is also conserved. 
In the case of spherical coordinates we have ${\bf v}=
\dot r\,dr+r^2\,\dot \vartheta\,d\vartheta+
r^2\sin^2{\vartheta}\dot\varphi\,d\varphi$, 
$v^2=g^{\mu\nu}v_\mu v_\nu=\dot r^2+r^2\dot\vartheta^2+
r^2\sin^2{\vartheta}\dot\varphi^2=v_0^2=const$, ${\bf A}{\bf v}=
g^{\varphi\varphi}A_\varphi v_\varphi=
(br+B)\dot\varphi$ with ${\bf A}$ from (8) and 
in accordance with (13) we obtain 
$$\mu(\ddot r-r\sin^2{\vartheta}\dot\varphi^2-
r\dot\vartheta^2)=\,\dot\varphi\,,\eqno
(14)$$
$$\frac{d}{dt}\left(r^2\dot\vartheta\right)-
r^2\dot\varphi^2\sin{\vartheta}\cos{\vartheta}
=0\,,\eqno(15)$$
$$\mu\frac{d}{dt}
\left(r^2\dot\varphi\sin^2{\vartheta}\right)
=-\,\dot r\>\eqno(16)$$
with dimensionless parameter $\mu=E/(qb)$. 
In the case of Cartesian coordinates we have 
${\bf v}=\dot x\,dx+\dot y\,dy+\dot z\,dz$, $v^2={\dot x}^2+
{\dot y}^2+{\dot z}^2=v_0^2=const$, ${\bf A}{\bf v}=
-\dot x\frac{(br+B)y}{x^2+y^2}+\dot y\frac{(br+B)x}{x^2+y^2}$ 
with ${\bf A}$ from (8)--(9) and (13) gives rise to
$$\mu\ddot x=\frac{1}{r}\left(\dot y+\dot z\frac{yz}{x^2+y^2}\right) , 
\eqno(17)$$
$$\mu\ddot y=-\frac{1}{r}\left(\dot x+\dot z\frac{xz}{x^2+y^2}\right) , 
\eqno(18)$$
$$\mu\ddot z=\frac{z}{r(x^2+y^2)}\left({x}\dot y-{y}\dot x\right)\>. 
\eqno(19)$$
Also we should add the initial conditions to (14)--(16) and (17)--(19). 
Namely, putting an initial moment of time $t_0=0$ for simplicity, we 
have, respectively, 
$$r(0)=r_0, \vartheta(0)=
\vartheta_0, \varphi(0)=\varphi_0, \dot r(0)=\dot{r}_0, \dot\vartheta(0)=
\dot{\vartheta}_0, \dot\varphi(0)=\dot{\varphi}_0\>. \eqno(20)  $$
or
$$x(0)=x_0, y(0)=y_0, z(0)=z_0, \dot x(0)=\dot{x}_0, 
\dot y(0)=\dot{y}_0, \dot z(0)=\dot{z}_0\>. \eqno(21)$$   


%
%
 \section{General considerations}
\subsection{Magnetic field lines}
 Let us above all find out how the magnetic field lines look for the field of 
(11). According to a general prescription (see, e.g., Ref. \cite{Car}) we should 
determine integral curves for differential system 
$$ \frac{dx}{H_x}=\frac{dy}{H_y}=\frac{dz}{H_z}\>, \eqno(22) $$
which can be made if finding the first integrals for it, i.e., such functions 
$\psi$ that satisfy the partial differential equation 
$$\frac{\partial\psi}{\partial x}H_x+\frac{\partial\psi}{\partial y}H_y+
\frac{\partial\psi}{\partial z}H_z =0\>.  \eqno(23)$$
Then, as is not complicated to check, the system (22) has two independent 
first integrals, namely $y=C_1x$, $x^2+y^2+z^2=C_2^2$ with constants $C_{1,2}$. That is, 
integral surfaces are planes and spheres and, as a result, integral curves are 
circles. For example, the plane $y=0$ is integral surface and equations of 
field lines are $x^2+z^2=C_2^2$ (see Fig. 1). 
\begin{figure*}
\vspace{0cm}
\caption{Magnetic field lines of the confining magnetic field}
\end{figure*}
\subsection{The confining properties}
We can note that 
$$\frac{d^2 }{dt^2}r^2=\frac{d}{dt}(2r\dot{r})=
2\frac{d }{dt}(x\dot{x}+y\dot{y}+z\dot{z})=
2[v^2+(x\ddot{x}+y\ddot{y}+z\ddot{z})]\eqno(24) $$
with $v^2=\dot{x}^2+\dot{y}^2+\dot{z}^2$. Multiplying (17), (18), (19) by 
$x,y,z$, respectively, and adding the results, we get 
$$\mu(x\ddot{x}+y\ddot{y}+z\ddot{z})=\frac{x\dot{y}-y\dot{x}}{r}
\left(1+\frac{z^2}{x^2+y^2}\right)\,,\>\mu=\frac{E}{qb}\>. \eqno(25)$$
To calculate $x\dot{y}-y\dot{x}$ we notice that
$\frac{d}{dt}(x\dot{y}-y\dot{x})=x\ddot{y}-y\ddot{x}$ and replacing 
$\ddot{x}, \ddot{y}$ according to (17) and (18), conformably, we shall have 
$\mu(x\ddot{y}-y\ddot{x})=-\frac{1}{r}(x\dot{x}+y\dot{y}+z\dot{z})=-\dot{r}$ 
wherefrom
$$\mu(x\dot{y}-y\dot{x})=-r+A_0 \>,  \eqno(26)$$ 
where a constant $A_0$ can be found from initial conditions (21) when 
considering (26) at $t=0$ so $A_0=\mu(x_0\dot{y}_0-y_0\dot{x}_0)+r_0$ and 
$$ \mu(x\dot{y}-y\dot{x})=-\sqrt{x^2+y^2+z^2}+\mu(x_0\dot{y}_0-y_0\dot{x}_0)
+r_0\>.\eqno(27)$$
We can consider $A_0\ge0$ which always holds true for the strong enough field 
when $|b|\to\infty$ and, consequently, $\mu\to0$. Then, considering (27) on  
$z$-axis where $x=y=0$, we obtain $|z|=A_0$ which signifies that particle 
trajectory can reach $z$-axis only at $z=\pm A_0$. It should be recalled that 
according to Sec. 2 the $z$-axis is forbidden for motion of a particle so if 
for the particle $|z|=A_0$ at some moment then after it one should consider 
the motion to be finished and the particle vanished. Now, using (26), we can 
rewrite (25) as 
$$\mu^2(x\ddot{x}+y\ddot{y}+z\ddot{z})=\frac{A_0-r}{r}
\left(1+\frac{z^2}{x^2+y^2}\right)=\frac{A_0-r}{r}(1+\cot^2{\vartheta})   
\eqno(28)$$
with spherical coordinate $\vartheta$. At last, with the help of (28) we 
derive from (24)
$$\frac{d^2 }{dt^2}r^2=\frac{d}{dt}(2r\dot{r})=2\left[v^2+
\frac{A_0-r}{r\mu^2}(1+\cot^2{\vartheta})\right]\>.   
\eqno(29)$$
At $r\le A_0$ from here it follows
$$\dot{r}=\frac{1}{r}\int\left[v^2+
\frac{A_0-r}{r\mu^2}(1+\cot^2{\vartheta})\right]dt>
\frac{1}{A_0}\int v^2dt>0\>, \eqno(30)$$
which signifies that $r$ is increasing. But if $r\ge2A_0$, i.e., 
$r-A_0\ge A_0$, then $A_0/r\le1/2$ and from (30) we gain 
$$\dot{r}=\frac{1}{r}\int\left[v^2-\frac{1}{\mu^2}+\frac{A_0}{r\mu^2}-
\frac{r-A_0}{r\mu^2}\cot^2{\vartheta}\right]dt<\frac{1}{2A_0}
\int \left(v^2-\frac{1}{2\mu^2}\right)dt<0\>, \eqno(31)$$
provided that
$$v<\frac{1}{\sqrt{2}|\mu|}=\frac{|qb|}{\sqrt{2}E}\>,\eqno(32)$$
i.e., $r$ is decreasing. It should be emphasized that for sufficiently strong 
field ($|b|\to\infty$) the condition (32) will always be fulfilled for the given 
$E$. Besides, under this situation, $A_0=\mu(x_0\dot{y}_0-y_0\dot{x}_0)+
r_0\sim r_0$ and we can see that spherical coordinate $r$ never tends to 
infinity and oscillates near the initial value $r_0$. Inasmuch as, as said 
above, $r=r_0$, $\varphi=\varphi_0$ is a magnetic field line, then we can say 
that the particle 
trajectory oscillates near the magnetic field line defined by initial 
conditions. 
In other words, we get the full confinement of charged particle in the 
magnetic field under discussion which is sketched out in Fig. 2. 
\begin{figure}
\vspace{0cm}
\caption{Behaviour of a charged particle in the confining magnetic field}
\end{figure}

\section{Asymptotical expansions}
To illustrate the general properties decribed in Section 3 it should be noted 
that the system (14)--(16) seems to be insoluble in an explicit form. But let 
us try to obtain an asymptotical solution of it in the form of expansions in 
the dimensionless parameter $\mu=E/(qb)$ in the strong field limit when $b\to\infty$, i.e., 
$\mu\to0$. For 
this aim we can notice that the angle $\varphi$ for Lagrangian $L$ of (12) is 
the so-called cyclic coordinate, i.e. $L$ does not depend on $\varphi$. Then in 
accordance with the Lagrange approach we have an integral of motion [see (13)] 
in the form 
$$\frac{\partial L}{\partial \dot{\varphi}}=Er^2\dot\varphi\sin^2{\vartheta}+q(br+B)=
\alpha_\varphi=Er_0^2\dot\varphi_0\sin^2{\vartheta_0}+q(br_0+B)=const\>, 
\eqno(33) $$
with using the initial data of (20). From here it follows 
$$\dot\varphi=\frac{\nu-r}{\mu r^2\sin^2{\vartheta}}, 
\>\nu=r_0+\mu\,r_0^2\dot\varphi_0\sin^2{\vartheta_0}\>,
 \eqno(34)  $$
and it is not complicated to rewrite the system (14)--(16) in the form 
$$\mu^2\frac{dp}{dt}=\mu^2\,\frac{s^2}{r^3}+
\frac{\nu(\nu-r)}{r^3\sin^2{\vartheta}} \>,\eqno(35)$$
$$\mu^2\frac{ds}{dt}=\frac{(\nu-r)^2\cos{\vartheta}}{r^2\sin^3{\vartheta}} 
\>,\eqno(36)$$
$$ \dot r=p \>,\eqno(37)$$
$$ \dot\vartheta=\frac{s}{r^2} \>\eqno(38)$$
with $ s=r^2\dot\vartheta$. 
\subsection{Expressions for $r$ and $\vartheta$}
Further we seek for $r$, $\vartheta$, $s$, $p$ in 
the form
$$r=\bar{r}_0+\mu\,\bar{r}_1+\mu^2\,\bar{r}_2+O(\mu^3) \>, 
\vartheta=\bar{\vartheta}_0+\mu\,\bar{\vartheta}_1+O(\mu^2) \>,$$
$$s=\bar{s}_0+\mu\,\bar{s}_1+O(\mu^2) \>, 
p=\bar{p}_0+\mu\,\bar{p}_1+O(\mu^2) \>,\eqno(39) $$
where $\bar{r}$, $\bar{r}_1$, $\bar{r}_2$, $\bar{\vartheta}_0$, 
$\bar{\vartheta}_1$, $\bar{s}_0$, $\bar{s}_1$, $\bar{p}_0$, 
$\bar{p}_1$ are some functions of $t$.    
Now, expanding the right-hand side of (35) in $\mu$, we obtain
$$\mu^2(\dot{\bar{p}}_0+\mu\,\dot{\bar{p}}_1)+O(\mu^4)=
\frac{r_0(r_0-\bar{r}_0)}
{\bar{r}_0^3\sin^2{\bar{\vartheta}_0}}+O(\mu)\>,$$
so we should have $\bar{r}_0=r_0$. 

Then the following terms of expansion for the 
right-hand side (35) give rise to relation 
$$\mu^2(\dot{\bar{p}}_0+\mu\,\dot{\bar{p}}_1)+O(\mu^4)=
\frac{r_0^2\dot{\varphi}_0\sin^2{\vartheta_0}-\bar{r}_1}
{\bar{r}_0^2\sin^2{\bar{\vartheta}_0}}\mu+O(\mu^2)\>,$$
which yields $\bar{r}_1=r_0^2\dot{\varphi}_0\sin^2{\vartheta}_0=const$. 
In this situation the new terms of expansion for 
the right-hand side (35) lead to
$$\mu^2(\dot{\bar{p}}_0+\mu\,\dot{\bar{p}}_1)+O(\mu^4)=
\left(\frac{\bar{s}_0^2}{r_0^3}-\frac{\bar{r}_2}
{{r}_0^2\sin^2{\bar{\vartheta}_0}}\right)\mu^2+O(\mu^3)\>$$ 
and it should be  
$$\dot{\bar{p}}_0=\frac{\bar{s}_0^2}
{{r}_0^3}-\frac{\bar{r}_2}{{r}_0^2\sin^2{\bar{\vartheta}_0}}\>.\eqno(40) $$  
Let us now pass on to the equation (36) where the conforming expansion with the 
help of (39) yields  
$$\mu^2(\dot{\bar{s}}_0+\mu\,\dot{\bar{s}}_1)+O(\mu^4)=
\frac{\bar{r}_2^2\cos{\bar{\vartheta}_0}}
{{r}_0^2\sin^3{\bar{\vartheta}_0}}\,\mu^4+O(\mu^5)\>. $$
From here we have $\dot{\bar{s}}_0=0$ and, consequently, 
$\bar{s}_0=C_0=const$. Accordingly 
the equation (37) gives rise to 
$\dot{\bar{r}}_0+\mu\,\dot{\bar{r}}_1+O(\mu^2)=
\bar{p}_0+\mu\,\bar{p}_1+O(\mu^2)$ which entails $\bar{p}_0=\dot{\bar{r}}_0=0$, 
$\bar{p}_1=\dot{\bar{r}}_1=0$ since we have above obtained that $\bar{r}_0=r_0$, 
$\bar{r}_1=r_0^2\dot{\varphi}_0\sin^2{\vartheta}_0=const$. In the circumstances 
the relation (40) gives 
$$\bar{r}_2=\frac{C_0^2}{r_0}\sin^2{\bar{\vartheta}_0}\>. \eqno(41)$$

At last, in a similar way from (38) we can obtain the relation 
$$\dot{\bar{\vartheta}}_0+\mu\,\dot{\bar{\vartheta}}_1 +O(\mu^2) =
\frac{\bar{s}_0}{r_0^2}+O(\mu)\>,  $$
which entails $\dot{\bar{\vartheta}}_0=\frac{\bar{s}_0}{r_0^2}=
\frac{C_0}{r_0^2}$ and, as a result,  
$\bar{\vartheta}_0=\frac{C_0}{r_0^2}t+C_1$ with 
some constant $C_{1}$. Then, taking into account (39) and (41), we finally 
have
$$r=r_0+\mu\,r_0^2\dot{\varphi}_0\sin^2{\vartheta}_0+
\mu^2\,\frac{C_0^2}{r_0}\sin^2{\left(\frac{C_0}{r_0^2}t+C_1\right)}+ 
O(\mu^3), \vartheta=\frac{C_0}{r_0^2}t+C_1+O(\mu)\>. \eqno(42)$$
\subsection{Expression for $\varphi$}
When searching for $\varphi$ in the form $\varphi=\bar{\varphi}_0+
\mu\,\bar{\varphi}_1+O(\mu^2)$ we shall, according to (34), obtain 
$$ \mu(\dot{\bar{\varphi}}_0+\mu\,\dot{\bar{\varphi}}_1)+O(\mu^3)= 
-\frac{\bar{r}_2}
{{r}_0^2\sin^2{\bar{\vartheta}_0}}\,\mu^2+O(\mu^3)\>,  $$
wherefrom, with the help of (41), we get 
$\dot{\bar{\varphi}}_0=0$, $\dot{\bar{\varphi}}_1=
-\frac{C_0^2}{r_0^3} $ and, consequently, $\bar{\varphi}_0=C_2$, 
$\bar{\varphi}_1=-\frac{C_0^2}{{r}_0^3}t+C_3 $ with some constants $C_{2,3}$. 
So finally 
$$\varphi=C_2+\left(C_3-\frac{C_0^2}{{r}_0^3}t\right)\mu+O(\mu^2)\>.
\eqno(43)$$ 
\subsection{Determination of constants}
At $\mu\to0$ with taking (20), (42) and (43) into account we find 
$C_0\approx r_0^2\dot{\vartheta}_0$, $C_1\approx\vartheta_0$, 
$C_2\approx\varphi_0$. This eventually leads to the final expressions
$$r\approx r_0+\mu\,r_0^2\dot{\varphi}_0\sin^2{\vartheta}_0+
\mu^2\,{r_0}^3\dot{\vartheta}_0^2\sin^2{(\dot{\vartheta}_0 t+
\vartheta_0)}+ O(\mu^3), 
\vartheta\approx\vartheta_0+\dot{\vartheta}_0\,t+O(\mu) \>, $$
$$\varphi\approx \varphi_0+(C_3-r_0\dot{\vartheta}_0^2\,t)\,\mu+O(\mu^2)
                   \>, \eqno(44) $$
that confirm the general considerations of Section 3 (see also Fig. 2). 
\section{Numerical estimates}
\subsection{General estimates}
As is clear from (44), if we want a charged particle to be near magnetic field 
line $r=r_0$, $\varphi=\varphi_0$ defined by intitial conditions then we should 
impose the condition 
$|\mu\,{r_0}^2\dot{\varphi}_0\sin^2{\vartheta_0}|<<r_0$ which entails 
$$ |\mu\,{r_0}\dot{\varphi}_0\sin^2{\vartheta_0}|<<1\>,\,
\mu=\frac{E}{qb}\>.\eqno(45)$$
If defining the physical 
components of velocity in spherical coordinates by equality  
$v^2=(v_r^{ph})^2+(v_\vartheta^{ph})^2+(v_\varphi^{ph})^2=
g^{\mu\nu}v_\mu v_\nu=\dot r^2+r^2\dot\vartheta^2+
r^2\sin^2{\vartheta}\dot\varphi^2=v_0^2=\dot r_0^2+r_0^2\dot\vartheta_0^2+
r_0^2\sin^2{\vartheta_0}\dot\varphi_0^2=const$ then we get $v_r^{ph}=\dot r$, 
$v_\vartheta^{ph}=r\dot\vartheta$, $v_\varphi^{ph}=
r\sin{\vartheta}\dot\varphi$ and the condition (45) signifies 
that $|\mu\,(v_0)_{\varphi}^{ph}\sin{\vartheta_0}|\le|\mu\,v_0|<<1$. 

But, obviously, $E=m/\sqrt{1-v_0^2}$ so the condition $|\mu\,v_0|<<1$ can be 
rewritten as
$$ \frac{mv_0}{\sqrt{(1-v_0^2)4\pi\,N^2\alpha_{em}}}<<|b|   \eqno(46)$$
with $q=Ne$ and the electromagnetic coupling constant $\alpha_{em}=
e^2/(4\pi)\approx
1/137.036$ in the chosen system of units. 
\subsection{Deuteron in thermonuclear plasma}
For this case typical values $r_0\sim1$ m (see e.g. Ref. \cite{Plasma}) and at 
temperature of plasma $T\sim0.8625\times10^{-2}$ MeV ($10^8$ K) we have a mean 
thermal deuteron velocity $v_0\sim\sqrt{3T/m}\approx0.372\times10^{-2}$ with 
the deuteron rest energy $m=m_p+m_n-2.225\,{\rm MeV}\approx(938+939-2.225)$ 
MeV $=1874.775$ MeV, $N=1$ and (46) yields $|b|>>23.0$ MeV. Let us take 
$|b|=1$ GeV and, for the sake of simplicity, put $\vartheta_0=\pi/2$. Then in 
accordance with (A.17) the module of magnetic field strength near the particle 
trajectory will be 
$$H=\frac{|b|}{r_0\sin{\vartheta_0}}\sim0.197\times10^{-15}\,{\rm GeV}^2\sim
0.351\, {\rm T}\>, \eqno(47)$$
i.e., it is a quite accessible value under the laboratory conditions. It should 
be, however, noted that for the time $t$ of order 1 s necessary to confine 
plasma before thermonuclear reaction starts \cite{Plasma} the angle $\vartheta$ 
can get increase 
$\dot{\vartheta}_0\,t=(v_0)^{ph}_{\vartheta}t/r_0\sim v_0t/r_0 \approx10^{6}$ 
according to (44), i.e. promptness of particle along the magnetic field line 
$r=r_0$, $\varphi=\varphi_0$ (see Fig. 2) will approximately be equal to 
$N_0=10^{6}/(2\pi)\approx 1.77\times10^5$, i.e., the particle will repeatedly 
cross $z$-axis which is impossible since the $z$-axis is forbidden for 
motion, as was mentioned in Sec. 2 and 3. 
But if $(v_0)^{ph}_{\vartheta}\to0$ 
then $N_0\to0$ as well. We can draw the conclusion that deuteron rushing in 
to the field with $(v_0)^{ph}_{\vartheta}\approx0$ will remain near its initial 
position during the time $t=1$ s. 
                                          
\subsection{Quarks in pions}
We may with minor reservations try applying the results obtained also to quarks 
within hadrons, e.g., within charged pions $\pi^{\pm}$. In this case quarks 
are moving in the classical confining SU(3)-gluonic field (3) but they are 
quntum objects described by the wave functions - the modulo square integrable 
solutions of the Dirac equation in the field (3) (for more details see
Refs. \cite{{Gon06},{Gon07a},{Gon08b}}). Let us, however, look at what the classical 
estimate (46) can give for quarks where, obviously, electric charge should be 
replaced by colour one and $\alpha_{em}$ by $\alpha_{s}$, the strong coupling 
constant. We can use the fact \cite{{Gon06},{Gon07a},{Gon08b}} that the colour magnetic field between 
quarks can be characterized by an effective colour strength 
$H=b/(r\sin{\vartheta})$ with $b=\sqrt{b_1^2+b_1b_2+b_2^2}$ and $b_{1,2}$ from 
the solution (3) while $r$ stands for the 
distance between quarks. Then (46) allow us to introduce quantity
$$b_0= \frac{mv_0}{\sqrt{(1-v_0^2)4\pi\,\alpha_{s}}} \eqno(48)$$
and for $u$-quark in $\pi^\pm$-mesons with $m=m_u\sim2.25$ MeV, $v_0\sim0.99$, 
$\alpha_{s}\sim0.485$ \cite{Gon08b} we obtain $b_0\approx22.68$ MeV which 
at the scale of pion $r_0\approx0.672$ fm entails 
$H=b_0/(r_0\sin{\vartheta})\sim0.666\times10^{-2}$ MeV$^2\sim 
0.119\times10^{14}$ T when $\vartheta=\pi/2$. In reality, from 
quantum considerations \cite{Gon08b} for $\pi^\pm$-mesons it follows 
$b_1=0.178915$ GeV, $b_2=-0.119290$ GeV so 
$b=\sqrt{b_1^2+b_1b_2+b_2^2}\approx0.157$ GeV $>b_0$ and at the scales of the 
meson under consideration we have \cite{Gon08b} $H\sim(10^{15}-10^{16})$ T. 
As a result, classical estimate corresponds to the quantum considerations. 

\section{Discussion and concluding remarks}
 It is useful to compare our results with the well-known case of 
motion of a charged particle in the homogeneous magnetic field 
(see, e.g., Ref. \cite{LL}) which is sketched out in Fig. 3.  
\begin{figure}
\vspace{0cm}
\caption{Motion of a charged particle in the homogeneous magnetic field}
\end{figure}
In the latter case the particle moves along helical curve with lead of helix 
$h=2\pi mv\cos{\alpha}/(qH\sqrt{1-v^2})$ and radius 
$R=mv\sin{\alpha}/(qH\sqrt{1-v^2})$. As a consequence, the homogeneous magnetic 
field does not give rise to the full confinement of the particle since the 
latter may go to infinity along the helical curve. Another matter is the case 
of the magnetic field (11). As we have seen above it provides the full 
confinement of any charged particle in case the field is strong enough: 
under any initial conditions the particle motion is accomplished 
within a finite region of space so that the particle trajectory is near 
magnetic field lines while the latter are compact manifolds (circles). This  
was explicitly demonstrated in Section 4 by obtaining asymptotical form of 
the motion under discussion.

Taking into account such remarkable properties of the magnetic field in 
question we may hope that it should find a number of applications, 
in particular, in thermonuclear plasma physics where the problem of confinement 
of plasma during a sufficiently long time has so far not solved in a 
satisfactory way \cite{Plasma}. But for it one should explore the possible ways 
of modeling the field (11) in laboratory conditions which is seemingly not   
such a simple task. One of possible physical realization is sketched out in 
Fig. 4 and is accomplished between two cone ferromagnetic pole 
pieces where the fact is used that magnetic field lines of ferromagnet are 
perpendicular to its surface. Of course, field lines inside the pole 
pieces (shown by dash lines) break off but if diameter $D\to0$ while the pole 
pieces are approaching, the whole construction tends to the the formal 
mathematical definition domain ${\mathbb{R}}^3\backslash \{z\}$. 
In this realization it is clear why particles will leave the field: they 
will just be absorbed by pole pieces when moving along the field line. So 
one needs to fit parameters (in particular, the values of module $H$) of the 
whole construction in such a way that a particle could remain on trajectory 
for a long enough time $t$ (e.g., for deuteron in thermonuclear plasma 
$t\sim1$ s according to Sec. 5) not reaching the pole pieces. 

\begin{figure}
\vspace{0cm}
\caption{One possible physical realization of the confining magnetic field}
\end{figure}

Finally, as has been said in Section 1, the main motivation of writing the 
given paper was to interpret the mechanism of quark confinement proposed in 
Refs.\cite{{Gon01},{Gon051},{Gon052}} in classical terms. As seems to us, we 
could to a certain degree do it.  

\begin{acknowledgements}
Yuri Goncharov is grateful to Prof. F. Nasredinov from Experimental Physics 
Department of the Sankt-Petersburg State Polytechnical University for 
discussion regarding the confining properties of the 
magnetic field of (11) and for pointing out the possible physical realization 
of it. 
\end{acknowledgements}

\section*{Appendix A}
In main body of the paper we employ the formulation of vector analysis on a 
region $\Omega$ in ${\mathbb R}^3$ from Ref. \cite{Po}. To our mind such a formulation is most 
convenient especially while working with using the arbitrary curvilinear 
coordinates. The essence of that formulation is in systematical use of both 
the Hodge star operator $\ast$ and the exterior differential $d$.  
\subsection*{Hodge star operator on ${\mathbb R}^3$ and Minkowski spacetime}
Let $M$ is a smooth manifold of dimension $n$ so we denote an 
algebra of smooth functions on $M$ as $F(M)$. In a standard way the spaces as
of smooth differential $p$-forms $\Lambda^p(M)$ ($0\le p\le n$) are defined 
over $M$ as modules over $F(M)$.  
If a (pseudo)riemannian metric $G=ds^2=g_{\mu\nu}dx^\mu\otimes dx^\nu$ is given 
on $M$ in local coordinates 
$x=(x^\mu)$ then $G$ can naturally be continued on spaces $\Lambda^p(M)$ 
by relation 
$$G(\alpha,\beta)={\rm det}\{G(\alpha_i,\beta_j)\}    \eqno(\mathrm A.1)$$
for $\alpha=\alpha_1\wedge\alpha_2...\wedge\alpha_p$, 
$\beta=\beta_1\wedge\beta_2...\wedge\beta_p$, where for 1-forms
$\alpha_i=\alpha_\mu^{(i)}dx^\mu$, $\beta_j=\beta_\nu^{(j)}dx^\nu$ we have 
$G(\alpha_i,\beta_j)=g^{\mu\nu}\alpha_\mu^{(i)}\beta_\nu^{(j)}$ with the 
Cartan's wedge (exterior) product $\wedge$. Under the circumstances the Hodge 
star operator $*$: $\Lambda^p(M)\to\Lambda^{n-p}(M)$ is defined for any 
$\alpha\in\Lambda^p(M)$ by
$$\alpha\wedge(*\alpha)=G(\alpha,\alpha)\omega_g\> \eqno(\mathrm A.2)$$
with the volume $n$-form 
$\omega_g=\sqrt{|\det(g_{\mu\nu})|}dx^1\wedge...dx^n$. 
For example, for 2-forms $F=F_{\mu\nu}dx^\mu\wedge dx^{\nu}$ we have
$$
F\wedge\ast F=(g^{\mu\alpha}g^{\nu\beta}-g^{\mu\beta}g^{\nu\alpha})
F_{\mu\nu}F_{\alpha\beta}
\sqrt{\delta}\,dx^1\wedge dx^2\cdots\wedge dx^n,\,\mu<\nu,\,\alpha<\beta 
\>\eqno(\mathrm A.3)
$$
with $\delta=|\det(g_{\mu\nu})|$.
If $s$ is the number 
of (-1) in a canonical presentation of quadratic form $G$ then two the most 
important properties of $*$ are 
$$ *^2=(-1)^{p(n-p)+s}\>,\eqno(\mathrm A.4)$$
$$ *(f_1\alpha_1+f_2\alpha_2)=f_1(*\alpha_1)+f_2(*\alpha_2)\> 
\eqno(\mathrm A.5)$$
for any $f_1, f_2 \in F(M)$, $\alpha_1, \alpha_2 \in\Lambda^p(M)$, i. e., 
$*$ is a $F(M)$-linear operator. By virtue of (A.5) for description 
of $*$-action in local coordinates it is enough to specify $*$-action on 
the basis elements of $\Lambda^p(M)$, i. e. on the forms 
$dx^{i_1}\wedge dx^{i_2}\wedge...\wedge dx^{i_p}$ with $i_1<i_2<...<i_p$ 
whose number is equal to $C_n^p=\frac{n!}{(n-p)!p!}$.

The most important case of $M$ in the given paper is a region $\Omega$ in 
the Euclidean space $\mathbb R^3$ 
with local Cartesian ($x,y,z$) or spherical 
($r, \vartheta, \varphi$) coordinates. 
The metric is given by either $ds^2=dx^2+dy^2+dz^2$ or 
$ds^2=dr^2+r^2(d\vartheta^2+\sin^2\vartheta d\varphi^2)$, and we shall obtain 
the $*$-action on the basis 
differential forms according to (A.2) in both the cases as 
$$\ast dx=dy\wedge dz,\> \ast dy=-dx\wedge dz,\> \ast dz=dx\wedge dy,\>$$
$$\ast(dx\wedge dy)=dz,\>\ast(dx\wedge dz)=-dy,
\>\ast(dy\wedge dz)=dx,\>
\ast(dx\wedge dy\wedge dz)=1,\>\eqno(\mathrm A.6)$$
$$\ast dr=r^2\sin{\vartheta}d\vartheta\wedge d\varphi,\>
\ast d\vartheta=-\sin{\vartheta}dr\wedge d\varphi,\>
\ast d\varphi=\frac{1}{\sin{\vartheta}}dr\wedge d\vartheta,\>$$
$$\ast(dr\wedge d\vartheta)=\sin{\vartheta}d\varphi,\>
\ast(dr\wedge d\varphi)=-\frac{1}{\sin{\vartheta}}d\vartheta,\>
\ast(d\vartheta\wedge d\varphi)=\frac{1}{r^2\sin{\vartheta}}dr,\>$$
$$\ast(dr\wedge d\vartheta\wedge d\varphi)=\frac{1}{r^2\sin{\vartheta}},\>
\eqno(\mathrm A.7)$$
so that on any $p$-form $\ast^2=1$, as should be in accordance with 
(A.4).

Let us also adduce for inquiring the conforming relations for the case of 
cylindrical coordinates $\rho, \varphi, z$, where $x=\rho\cos{\varphi}$, 
$y=\rho\sin{\varphi}$ and metric 
$ds^2=d\rho^2+\rho^2d\varphi^2+dz^2$. Then
$$\ast d\rho=\rho d\varphi\wedge dz,\>
\ast d\varphi=-\frac{1}{\rho}d\rho\wedge dz,\>
\ast dz=\rho d\rho\wedge d\varphi,\>$$
$$\ast(d\rho\wedge d\varphi)=\frac{1}{\rho}dz,\>
\ast(d\rho\wedge dz)=-\rho d\varphi\>,
\ast(d\varphi\wedge dz)=\frac{1}{\rho}d\rho,\>$$
$$\ast(d\rho\wedge d\varphi\wedge dz)=\frac{1}{\rho}\>,
                                     \eqno(\mathrm A.8)$$
where we can again see that $\ast^2=1$ on any $p$-form according to (A.4).  

Also the important case of $M$ is Minkowski spacetime with coordinates 
$t, r, \vartheta, \varphi$ and metric 
$ds^2=dt^2-dr^2-r^2(d\vartheta^2+\sin^2\vartheta d\varphi^2)$, where we have

$$\ast dt=r^2\sin{\vartheta}dr\wedge d\vartheta\wedge d\varphi,\>
\ast dr=r^2\sin{\vartheta}dt\wedge d\vartheta\wedge d\varphi,\>$$
$$\ast d\vartheta=-\sin{\vartheta}dt\wedge dr\wedge d\varphi,\>
\ast d\varphi=\frac{1}{\sin\vartheta}dt\wedge dr\wedge d\vartheta,\>$$
$$\ast(dt\wedge dr)=-r^2\sin\vartheta d\vartheta\wedge d\varphi\>,
\ast(dt\wedge d\vartheta)=\sin\vartheta dr\wedge d\varphi\>,$$
$$\ast(dt\wedge d\varphi)=-\frac{1}{\sin\vartheta}dr\wedge d\vartheta\>,
\ast(dr\wedge d\vartheta)=\sin\vartheta dt\wedge d\varphi\>,$$
$$\ast(dr\wedge d\varphi)=-\frac{1}{\sin\vartheta}dt\wedge d\vartheta\>,
\ast(d\vartheta\wedge d\varphi)=\frac{1}{r^2\sin\vartheta}dt\wedge dr\>,$$
$$\ast(dt\wedge dr\wedge d\vartheta)=\sin\vartheta d\varphi\>,
\ast(dt\wedge dr\wedge d\varphi)=-\frac{1}{\sin{\vartheta}}d\vartheta,\>$$
$$\ast(dt\wedge d\vartheta\wedge d\varphi)=\frac{1}{r^2\sin{\vartheta}}dr,\>
\ast(dr\wedge d\vartheta\wedge d\varphi)=
\frac{1}{r^2\sin{\vartheta}}dt,\>
\eqno(\mathrm A.9)$$
so that on 2-forms $\ast^2=-1$, as should be in accordance with 
(A.4). More details about the Hodge star operator can be found in \cite{Bes87}. 

At last, it should be noted that all the above is easily over linearity 
continued on the matrix-valued differential forms (see, e. g., 
Ref. \cite{Car}), i. e., on the arbitrary 
linear combinations of forms 
$a_{i_1i_2...i_p}dx^{i_1}\wedge dx^{i_2}\wedge...\wedge dx^{i_p}$, 
where coefficients $a_{i_1i_2...i_p}$ belong to some space of matrices $V$, 
for example, a SU($3$)-Lie algebra. But now the 
Cartan's wedge (exterior) product $\wedge$ should be understood as  
product of matrices with elements consisting of usual (scalar) differential 
forms. In the SU(3)-case, if $T_a$ are 
matrices of 
generators of the SU($3$)-Lie algebra in $3$-dimensional representation, we 
continue the above scalar product $G$ on the SU($3$)-Lie algebra valued 1-forms 
$A=A^a_\mu T_adx^\mu$ and $B=B^b_\nu T_bdx^\nu$ by the relation
$$G(A,B)=g^{\mu\nu}A^a_\mu B^b_\nu{\rm Tr}(T_aT_b)\,,\eqno(\mathrm A.10)$$
where Tr signifies the trace of a matrix, 
and, on linearity with the help of (A.1), $G$ can be continued 
over any SU($3$)-Lie algebra valued forms. Such a matrix-valued generalization 
of $\ast$-operator is extremely useful when exploring solutions of the 
Yang-Mills (and Maxwell) equations in Minkowski spacetime and was 
systematically employed in Refs. \cite{{Gon01},{Gon051},{Gon052}}. 

\subsection*{Operations of vector analysis in terms of $\ast$ and $d$} 
The basic property of the exterior differential $d$ is the same 
form in the arbitrary curvilinear coordinates $x_i$ on $\Omega$. Namely, 
$d=\sum\limits_i\partial_{x_i}dx_i$ with 
$\partial_{x_i}=\partial/\partial x_i$. 
For example, in Cartesian and spherical coordinates 
we have, respectively, $d=\partial_xdx+\partial_ydy+\partial_zdz$ or 
$d=\partial_r dr+\partial_\vartheta d\vartheta+\partial_\varphi d\varphi$. 

Passing on now to the vector analysis on $\Omega$, we should note that 
it is usually formulated in Cartesian coordinates, where all main operations (
divergence, curl operator and so on) look in the simplest form which makes 
difficulties when transfering to the arbitrary curvilinear coordinates. We 
can, however, simplify the situation if noting that there is 
one-to-one correspondence between any vector field 
${\bf a}=a_x{\bf i}+a_y{\bf j}+a_z{\bf k}$ with $a_{x,y,z}=a_{x,y,z}(x,y,z)$  
and 1-form $a_xdx+a_ydy+a_zdz$ so that the latter will be denoted by the same 
notation ${\bf a}$ in what follows. When describing vector fields by 1-forms, 
however, we at once gain a number of advantages. Indeed, according to the 
standard rules \cite{Car} it is easy to get an expression of ${\bf a}$ in 
the arbitrary curvilinear coordinates $x^i$, $i=1,2,3$, if knowing $x,y,z$ as 
the functions $x^i$. This is done by replacing 
$dx=(\partial x/\partial x^i)dx^i$, $dy=(\partial y/\partial x^i)dx^i$, 
$dz=(\partial z/\partial x^i)dx^i$, $x,y,z=x,y,z(x^i)$ in the expression 
of ${\bf a}$ in Cartesian 
coordinates and we at one blow obtain the components of ${\bf a}$ in the 
given curvilinear coordinates -- those are just the coefficients at the 
corresponding $dx^i$. Further, the basic products of vectors, scalar and 
vectorial ones, are now simply written through the conforming differential 
forms. Namely, scalar product ${\bf (a,b)}\equiv{\bf ab}=g^{\mu\nu}a_\mu b_\nu$, 
vectorial product ${\bf a}\times {\bf b}=\ast({\bf a}\wedge{\bf b})$, where 
metric coefficients $g_{\mu\nu}$ and the Hodge star operator $\ast$ are defined by 
the expression $ds^2$ in the given curvilinear coordinates [see, e.g., 
(A.6)--(A.8)]. 

At last, we have the standard de Rham complex \cite{{Po},{Bes87}}
$$F(\Omega)=\Lambda^0\stackrel{d}{\longrightarrow}{\Lambda^1}
\stackrel{d}{\longrightarrow}{\Lambda^2}\stackrel{d}{\longrightarrow}
{\Lambda^3}\stackrel{d}\longrightarrow{0}\>,\eqno(\mathrm A.11)$$
which enables us to write down the main operations of vector analysis as 
$${\mathrm{grad}}\, f=d f=\frac{\partial f}{\partial x^i}dx^i \>,
{\mathrm{rot}}\, {\bf a}=\ast(d{\bf a})\>,
{\mathrm{div}}\, {\bf a}=\ast(d\ast{\bf a})
\eqno(\mathrm A.12)$$
with arbitrary function $f\in F(\Omega)$, so that, in virtue of the famous 
property $d^2=0$ for operator $d$, we automatically obtain the identities 
${\mathrm{rot}}\, {\mathrm{grad}} f\equiv 0 $, ${\mathrm{div}}\, {\mathrm{rot}} 
\,{\bf a}\equiv 0 $, provided that the first and second de Rham cohomology 
groups of $\Omega$ are equal to zero: $H^1\Omega=H^2\Omega=0$. It should be 
emphasized that relations (A.12) hold true for any curvilinear coordinates 
as soon as the expression of metric $ds^2$ is fixed in those coordinates and, 
accordingly, the Hodge star operator action is defined on any $p$-form. After 
computing with using the given curvilinear coordinates we can always return 
to Cartesian ones by replacing $dx^i=(\partial x^i/\partial x)dx+ 
(\partial x^i/\partial y)dy+(\partial x^i/\partial z)dz$, $dx\to{\bf i}$, 
$dy\to{\bf j}$, $dz\to{\bf k}$. 

Many relations of vector analysis obtained in Cartesian coordinates can easily 
be generalized to the arbitrary curvilinear coordinates within the formulation 
under consideration. For example, the identities
$$ {\mathrm {div}}({\bf a}\times{\bf b})=({\mathrm {rot}}\,{\bf a}){\bf b}-
{\bf a}({\mathrm {rot}}\,{\bf b})\>, \eqno(\mathrm A.13)$$
$$ {\mathrm {rot}}({\bf a}\times{\bf b})=({\bf b}\nabla){\bf a}-
({\bf a}\nabla){\bf b}+{\mathrm {div}}({\bf b}){\bf a}-
{\mathrm {div}}({\bf a}){\bf b}\> \eqno(\mathrm A.14)$$
acquire the form
$${\mathrm {div}}({\bf a}\times{\bf b})=\ast d({\bf a}\wedge{\bf b})\>,
\eqno(\mathrm A.15)$$
$$ {\mathrm {rot}}({\bf a}\times{\bf b})=\ast d\ast({\bf a}\wedge{\bf b})
\eqno(\mathrm A.16)$$
holding true for any curvilinear coordinates. On the other hand, if trying 
to use (A.13)--(A.14) 
within framework of the standard formulation, e.g., in spherical coordinates,
then this will lead to the perfectly bulky expressions. At the same time, the 
right-hand sides of (A.15)--(A.16) are easily computed for concrete 
${\bf a}$ and ${\bf b}$ in any curvilinear coordinates along the lines above. 

\subsection*{The confining magnetic field}
To illustrate some of the above let us compute the strength ${\bf H}$ of the 
confining magnetic field of (8)--(9) in Cartesian coordinates. By definition we 
have ${\bf H}={\mathrm {rot}}\,{\bf A}$ and in spherical coordinates 
${\bf A}=(br+B)d\varphi$. Then in accordance with (A.12) and (A.7) 
${\bf H}=\ast(d{\bf A})=\ast d[(br+B)d\varphi]=\ast(bdr\wedge d\varphi)=
-\frac{b}{\sin{\vartheta}}d\vartheta=H_\vartheta d\vartheta$ which entails 
the module $H=\sqrt{g^{\mu\nu}H_\mu H_\nu}=
\sqrt{g^{\vartheta\vartheta}H^2_\vartheta}=\frac{|b|}{r\sin{\vartheta}}$. 

On the other hand, $\vartheta=\arccos{\frac{z}{r}}=
\arccos{\frac{z}{\sqrt{x^2+y^2+z^2}}}$ wherefrom 
$$d\vartheta=
\frac{\partial \vartheta}{\partial x}dx+
\frac{\partial \vartheta}{\partial y}dy+
\frac{\partial \vartheta}{\partial z}dz=
\frac{xz}{r^2\sqrt{x^2+y^2}}dx+\frac{yz}{r^2\sqrt{x^2+y^2}}dy-
\frac{\sqrt{x^2+y^2}}{r^2}dz\>. $$
At last, $\sin{\vartheta}=\frac{\sqrt{x^2+y^2}}{r}$ and we get 
$${\bf H}= -\frac{b}{\sin{\vartheta}}d\vartheta=
-\frac{b}{r}\left(\frac{xz}{x^2+y^2}dx+\frac{yz}{x^2+y^2}dy-
dz\right) $$
or, using the above isomorphism $dx\Longleftrightarrow{\bf i}$, 
$dy\Longleftrightarrow{\bf j}$, $dz\Longleftrightarrow{\bf k}$, 
$${\bf H}= -\frac{b}{r}\left(\frac{xz}{x^2+y^2}{\bf i}+
\frac{yz}{x^2+y^2}{\bf j}-{\bf k}\right)\>,\eqno(\mathrm A.17) $$
so the module $H=\sqrt{H_x^2+H_y^2+H_z^2}=\frac{|b|}{\sqrt{x^2+y^2}}
=\frac{|b|}{r\sin{\vartheta}}$.  

\section*{Appendix B}
The facts adduced here have been obained in Refs. \cite{{Gon051},{Gon052}} and 
we concisely give them only for completeness of discussion in Section 2.

For the case of U(1)-group the Yang-Mills equations (1) turn into the second pair of 
Maxwell equations 
$$d\ast F= 0 \eqno(\mathrm B.1)$$ 
with $F=dA$, $A=A_\mu dx^\mu$. 
The most general ansatz for a spherically symmetric solution is $A=A_t(r)dt+A_r(r)dr+
A_\vartheta(r)d\vartheta+A_\varphi(r)d\varphi$. 

For the latter ansatz we have $F=dA=-\partial_rA_tdt\wedge dr+
\partial_rA_\vartheta dr\wedge d\vartheta+
\partial_rA_\varphi dr\wedge d\varphi$ 
for an arbitrary $A_r(r)$. Then, according to (A.9), we obtain 
$$\ast F= (r^2\sin{\vartheta})\partial_rA_td\vartheta\wedge d\varphi +
\sin{\vartheta}\partial_rA_\vartheta dt\wedge d\varphi-
\frac{1}{\sin{\vartheta}}\partial_rA_\varphi dt\wedge d\vartheta 
\eqno(\mathrm B.2)$$ 
which entails 
$$d\ast F= 
\sin{\vartheta}\partial_r(r^2\partial_rA_t)dr\wedge d\vartheta\wedge d\varphi-
\sin{\vartheta}\partial_r^2A_\vartheta dt\wedge dr\wedge d\varphi-$$
$$\cos{\vartheta}\partial_rA_\vartheta dt\wedge d\vartheta\wedge d\varphi+
\frac{1}{\sin{\vartheta}}\partial_r^2A_\varphi dt\wedge dr\wedge d\vartheta
\>,\eqno(\mathrm B.3)$$
wherefrom one can conclude that
$$\partial_r(r^2\partial_rA_t)=0,\>\partial^2_rA_\varphi=0\>,
\eqno(\mathrm B.4)$$
$$\partial^2_rA_\vartheta=\partial_rA_\vartheta=0\>.\eqno(\mathrm B.5)$$
and we draw the conclusion that 
$A_\vartheta=C_1$ with some constant $C_1$. But then the Lorentz 
condition (2) for the given ansatz gives rise to 
$$\sin{\vartheta}\partial_r(r^2A_r)+
\partial_\vartheta(\sin{\vartheta}A_\vartheta)=
\sin{\vartheta}\partial_r(r^2A_r)+
\partial_\vartheta(C_1\sin{\vartheta})=0, $$
or 
$$\partial_r(r^2A_r)+C_1\cot{\vartheta}=0, \eqno(\mathrm B.6)$$
which yields $A_r=-C_1\cot{\vartheta}/r+C_2/r^2$ with a constant $C_2$.  
But the confining solutions should be spherically symmetric and contain only 
the components which are 
Coulomb-like or linear in $r$, so one should put $C_1=C_2=0$. Consequently, 
the ansatz $A=A_t(r)dt+A_\varphi(r)d\varphi$ is the most general 
spherically symmetric one and then equations (B.4) give

$$ A_t =\frac{a}{r}+A \>, A_\varphi=br+B \>\eqno(\mathrm B.7)$$
with some constants $a, b, A, B$ parametrizing solutions which proves the 
uniqueness theorem of Section 2 for U(1)-group. Minor modification of the 
above considerations allows us to spread the proof to 
the Yang-Mills equations (1)
(for more details see Refs. \cite{{Gon051},{Gon052}}).



\includepdf[pages={1,}]{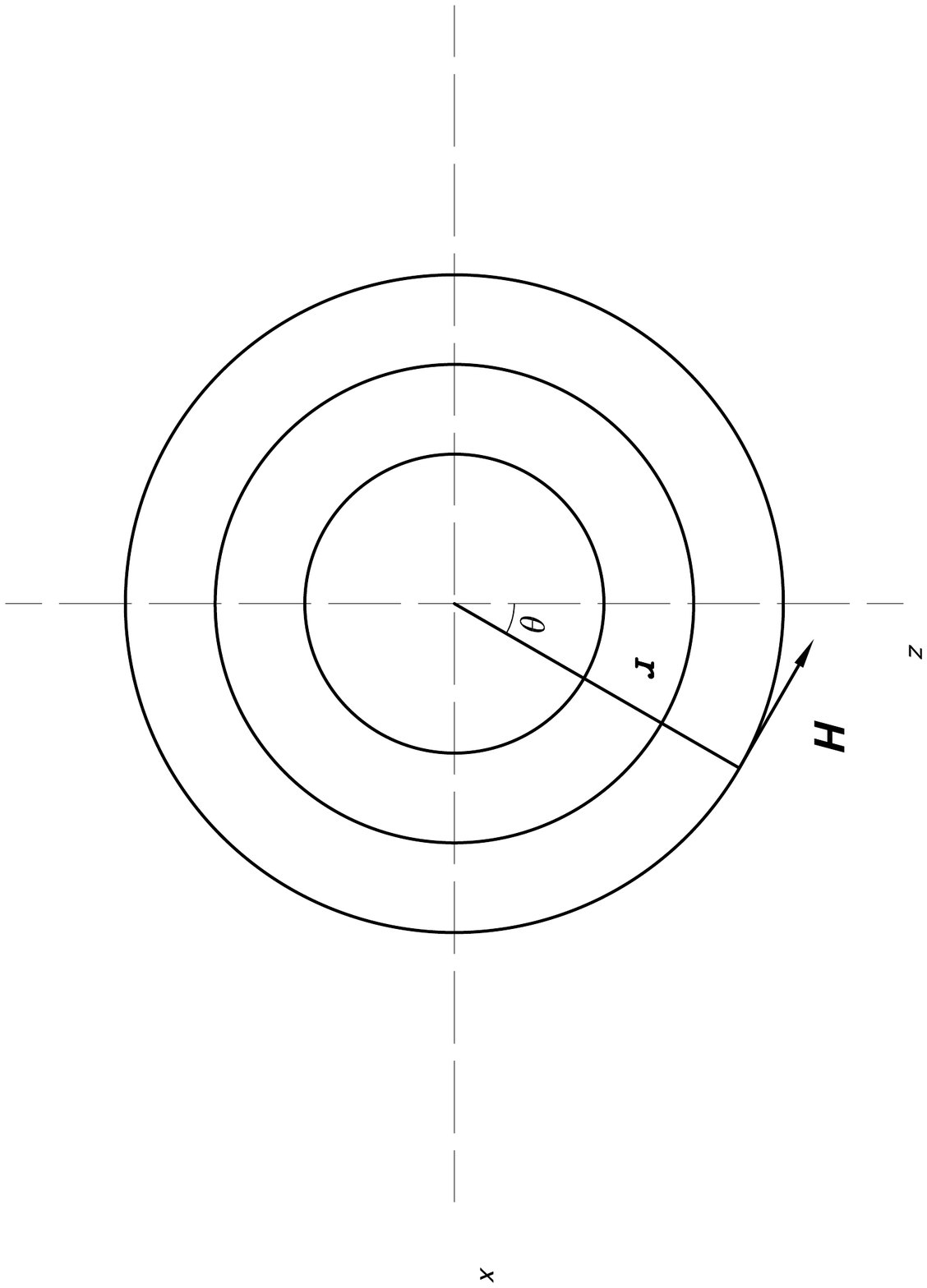}
\includepdf[pages={1,}]{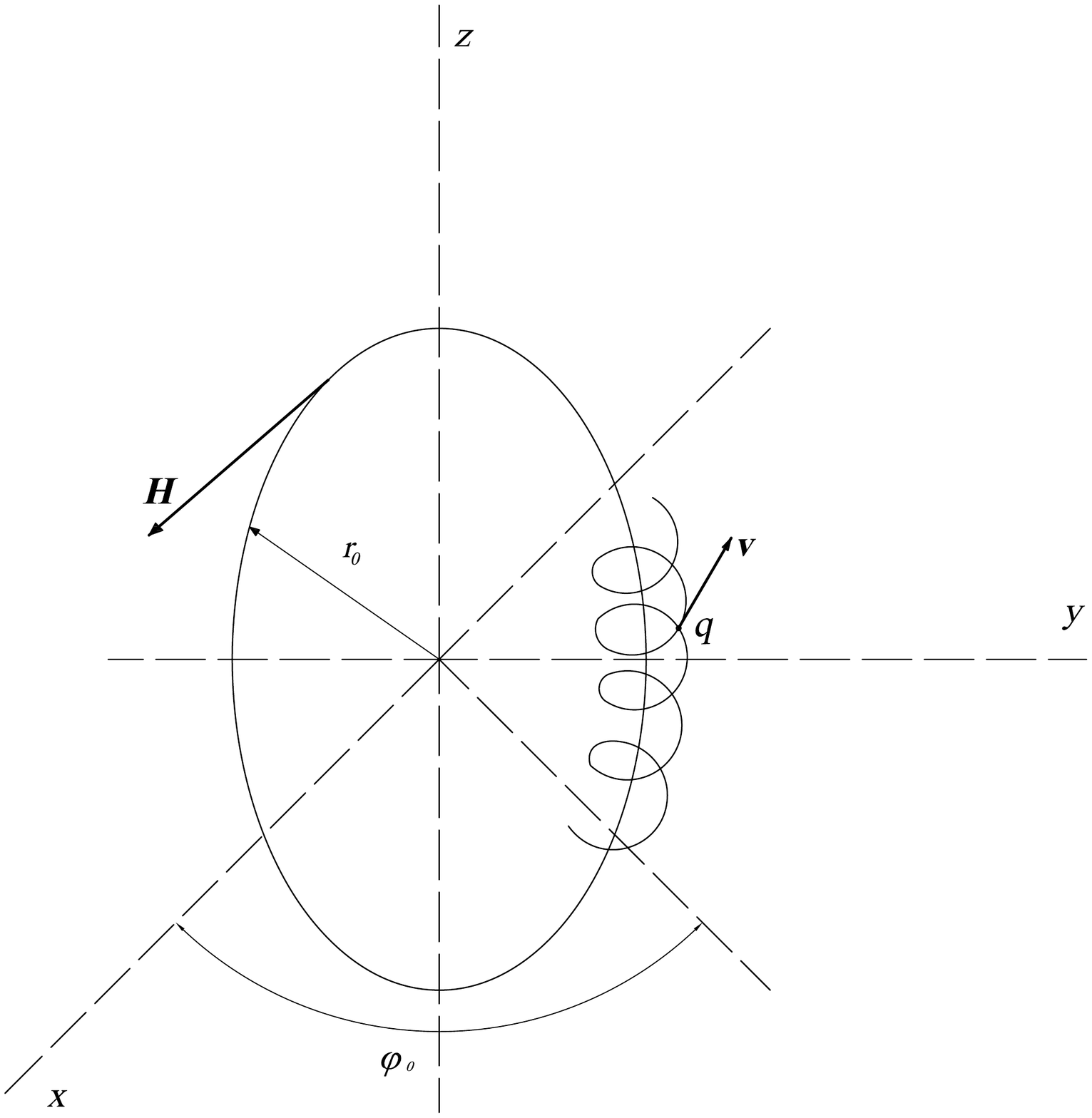}
\includepdf[pages={1,}]{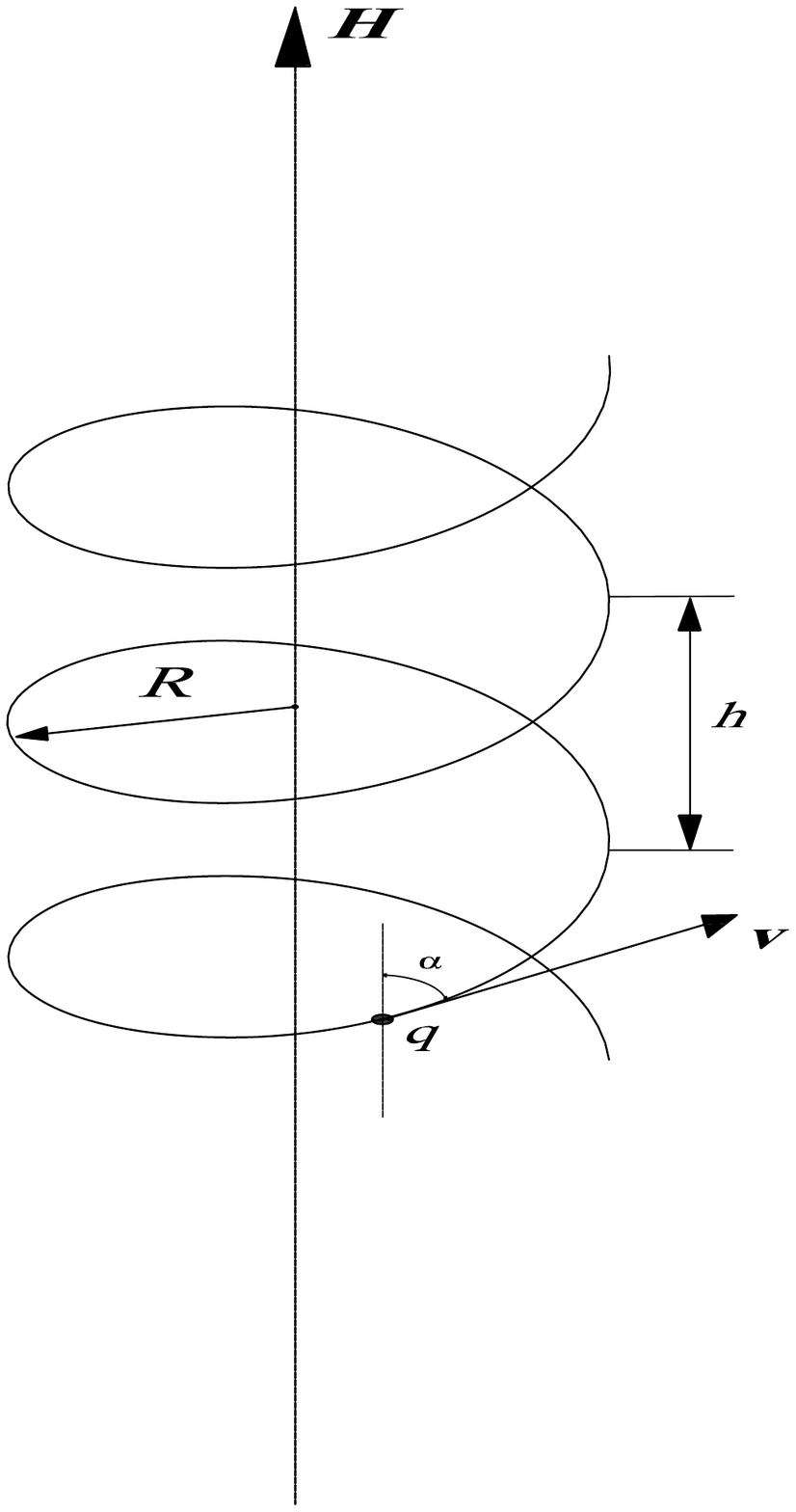}
\includepdf[pages={1,}]{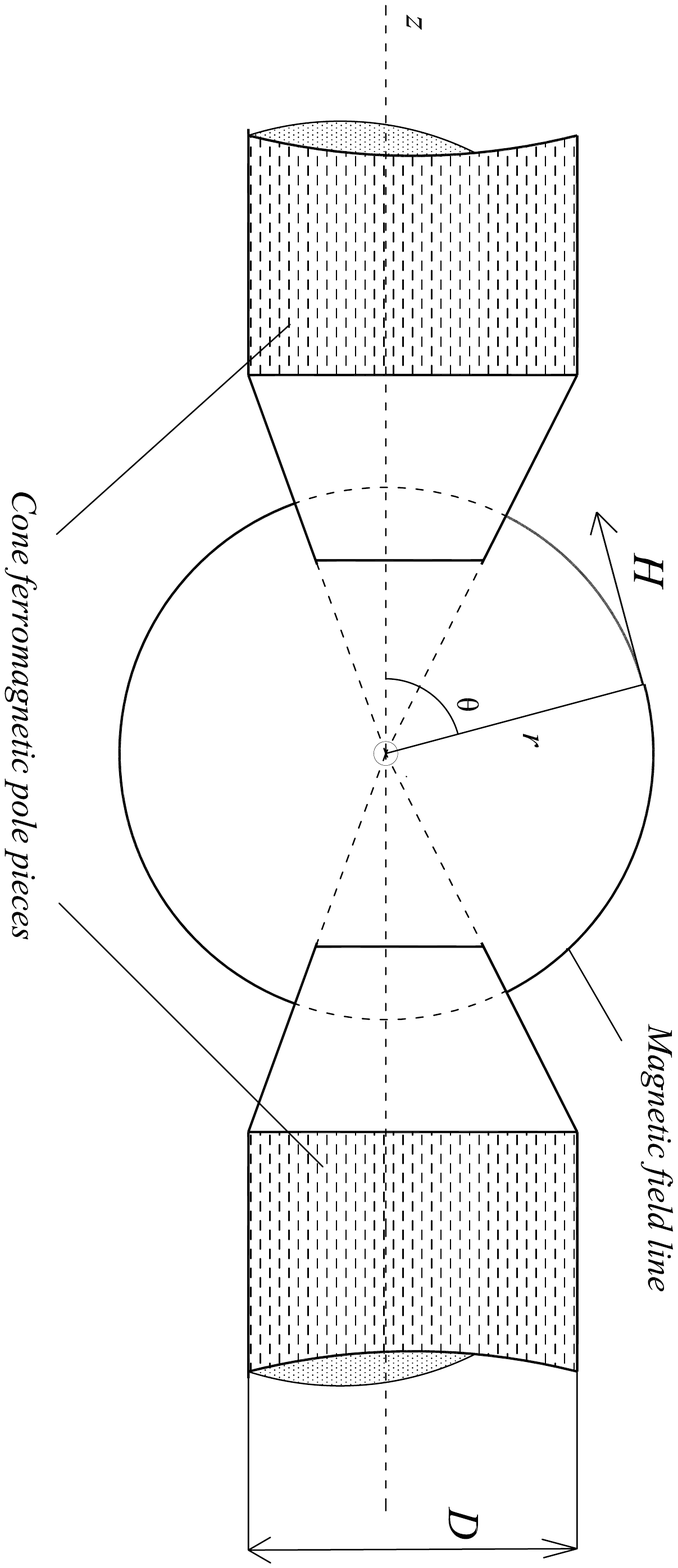}

\end{document}